\documentstyle[prb,aps,twocolumn]{revtex}

\begin{document}
\draft

\title{Relaxation times hierarchy in two-component quasiparticle gas}
\author{Alexander V. Zhukov}
\address{
Department of Theoretical Physics, Belgorod State University, 12
Studencheskaya str.,\\
308007 Belgorod, Russian Federation}

\date{July 8, 2001}
\maketitle

\begin{abstract}
A quasiparticle description of various condensed media is a very
popular tool in study of their transport and thermodynamic
properties. I present here a microscopic theory for the
description of diffusion processes in two-component gas of
quasiparticles with arbitrary dispersion law and statistics.
Particularly, I analyze the role of interaction within each
subsystem ({\it i.e.} between identical quasiparticles) in
relaxation of the whole system. The approach for solution of such
kinetic problem allows to study the most important limiting cases
and to clarify their physical sense. Classical results for
diffusion coefficient of light particles in a massive gas (Lorentz
model) and of massive particles in a light gas (Rayleigh model)
are obtained directly from the general solution without using
artificial approaches, as it was done earlier. This particularly
provide a possibility to generalize these popular models on
quasiparticle systems.
\end{abstract}

\section{Introduction}

It is well known that variety of properties of some condensed
media can be described by the interaction processes in
quasiparticle gases. These are, say, transverse and longitudinal
phonons in solids, phonons and magnons in magnetically ordered
materials, phonons and rotons in superfluid helium, conduction
electrons and holes in semiconductors {\it etc.} As a rule, to
study the dissipative properties of such systems investigators use
either classical kinetic theory in its simplest form or some
semi-intuitive models, which lead sometimes to quite ambiguous
results in case of quasiparticle systems. Here I present the
theory of diffusion processes in two-component quasiparticle
systems, which in general is independent on particular
quasiparticles statistics and dispersion law. Let me first briefly
review the present state of classical kinetic theory and analyze
its pitfalls in description of quantum quasiparticle systems of
condensed media.

 \par
The kinetic theory of gases in modern understanding takes its
beginning from pioneering work by Maxwell \cite{1}, in which he
has proved the law for distribution of velocities of molecules in
homogeneous equilibrium gas (so-called Maxwell velocities
distribution) and law of an equidistribution of average energy of
molecules in a mixture of gases. These his results were updated
and improved in further works devoted to the theory of
inhomogeneous gases (about a history of a problem see Ref. 2).
However as basis of all mathematical methods of the modern kinetic
theory it is necessary to consider apparently the basic works by
Boltzmann \cite{3}, in which both the H-theorem was proved and the
classical Boltzmann equation was introduced.

 \par
Boltzmann equation is the integro-differential equation describing
the collisional behaviour of rarefied gas. Until now it remains to
be a basis of the kinetic theory of gases and it appears to be
very fruitful not only for a research of classical gases, which
Boltzmann himself kept in mind, but - with an appropriate
generalization - for study of electron transport in solids and
plasma, transport of neutrons in nuclear reactors, phonon and
roton transport in superfluid liquids, and transport of a
radiation in atmospheres of stars and planets. For past 130 years
these researches have led to significant achievements both in new
areas, and in old one.

 \par
Generally, kinetic equation of the Boltzmann type (the equation
describing the evolution of a single--particle distribution
function in the phase space) represents the integro-differential
equation, which remarkable property is the nonlinearity of
collision term. Just this fact makes a main obstacle in
construction of methods for solution of the kinetic equation. The
monographies \cite{2,5} are devoted to a detail exposition of such
methods in case of classical gas systems.

 \par
In majority of experimental problems there is no necessity to use
the detail microscopic description of gas systems at a level of
distribution functions. As a rule, investigation of physical
processes in macroscopic systems is carried out at a less detail
level of hydrodynamic variables. Since these variables are
determined through the moments of a distribution function, then,
as a rule, the detail study of main moments of distribution
function appropriate to collisions invariants is required, but not
the distribution function itself. Thus, the connection between the
kinetic theory and hydrodynamics appears to be one of the main
problem. In particular, one of the main aspects of this problem is
the determination of transport coefficients, such as diffusion
coefficient, viscosity (first and second), thermal conductivity
appeared in equations of hydrodynamics of a viscous liquid
\cite{6}.

 \par
In spite of so long history of physical kinetics, today we have
rather small number of approaches to a solution of the kinetic
equations. All these approaches and methods were formed depending
on concrete problems, on which were directed. Most general methods
of research of non-equilibrium state of classical and quantum
gases were directed rather on demonstrating of a mathematical
resolvability (or insolubility) of certain basic problems in
principle, than on construction of the serviceable theories,
suitable for a solution of concrete physical problems.

 \par
The classical methods of a solution of the kinetic equations allow
to derive the kinetic coefficients as series expansions on an
infinite set of orthogonal polynomials. However, it appeared to be
very difficult to use these classical results for numerical
calculation and analysis of physical processes in real systems. In
particular, it is caused by impossibility of selection of the
contribution from different types of interactions to various
kinetic coefficients, while many physical systems behave
qualitatively differently for different ratio between a speed of a
relaxation inside each subsystem and between subsystems.

 \par
In classical gases with point-like interaction this problem is
less important because the speed of relaxation inside subsystems
unambiguously determined by their mass and concentration. In the
quantum case, when we talk about quasiparticles, the situation
becomes more complicated. The mechanism of interaction between
quasiparticles can be any independently on simple macroscopic
parameters. For example, in some cases such notation as mass
cannot be well defined at all (say, what is the mass of phonon?).

 \par
The mathematical theory of transport processes is most
sequentially advanced for mixtures of classical gases, the
evolution of which is described by a set of Boltzmann equations. A
basis of classical methods for solution of the Boltzmann equation
in case of one--component gas is the formal expansion of
distribution function in power series of some parameter $\sigma$
in the form $f=f^{(0)}+\sigma f^{(1)}+\sigma^2f^{(2)}+\dots$, so
that function $f^{(0)}$ corresponds to statistical equilibrium. In
this case parameter $\sigma$ is some scale factor for density, the
physical sense of which can be different depending on a particular
problem. As a rule, this parameter is formally considered to be
small, so that the solution of the kinetic equation represents a
problem of singular perturbation \cite{7}. The most successful
methods of a solution of the kinetic equation, such as the Hilbert
method and the Chapman--Enskog method \cite{2,5}, are based on
this principle. In spite of the successes of the Chapman--Enskog
method in the description of connection of the kinetic theory with
the equations of hydrodynamics (the Navier-Stokes equations appear
already in the first order in parameter $\sigma$), the explicit
expressions for kinetic coefficients have a rather complicated
form. Main defect of these expressions is that already in the
first order in parameter $\sigma$ their analysis becomes
practically impossible. The situation becomes more problematic in
the case of two--component gas. The infinite series of integral
brackets containing Sonin polynomials does not allow to select
explicitly the contributions from different types of interactions
in a system to various dissipative coefficients. This frequently
leads to necessity to use various ungrounded approximations, such
as the Chapman--Cowling approximation \cite{5} or Kihara
approximation \cite{2}.

 \par
At the same time, in spite of some successes \cite{8}, there is no
yet consistent mathematical theory for deriving the dissipative
coefficients in gases of quasiparticles. And, naturally, the
problem of distinguishability of contribution from interactions
between identical quasiparticles and between different subsystems
has not been solved yet. This problem has his own history. In
practice, when analyzing particular physical systems, many
physicists use some model approximations for collision integrals.
The most popular one is the so--called BGK--approximation
\cite{9}. In its simple form BGK--approximation leads sometimes to
quite confusing results. Really, let us have several relaxation
mechanisms in the system, and, therefore several characteristic
times depending on momenta or energies of quasiparticles. The
final observed quantities should be obtained by averaging this
times in some manner. In particular problems the following
question frequently appears: What must be averaged, {\it e.i.} the
time or the rate (inverse time)? and how to obtain the real
relaxation time, {\it i.e.} by summation of the times or by
summation of the rates? This uncertainty led to many confusing
situations. For example, more than for 15 years there were two
different theories for the mobility of two--dimensional electron
gas localized over the free surface of liquid helium \cite{10}.
The first theory \cite{11} assumes the mobility to depend on the
averaged characteristic time of electron--ripplon interaction
(ripplons are the quantized surface waves of liquid helium). This
theory well describes the experimental data for small electron
density \cite{12}. Another theory \cite{13} assumes the mobility
to be determined by the inverse averaged rate of the same
interaction. These theoretical results well fit another
experiments \cite{14} with large electron density. The problem of
relationship between these two results has naturally appeared.
Similar confusion was in the theory of dissipative processes in
superfluids \cite{15}. The analogous situation took place for some
time in theory of thermal conductivity in solids \cite{16}.

 \par
The aim of this paper is to present alternative approach for
solution of the system of linearized kinetic equations for
two--component gas of quasiparticles with arbitrary statistics and
dispersion law. The theory explicitly accounts the all types of
interactions in the system. This allows to analyze the
contribution of interaction between identical particles into the
relaxation of the whole system. I do not restrict myself here in
the frame of particular system. So, the results obtained here can
be applied to any quantum system, whose dissipative properties are
determined by the processes in two--component gas of
quasiparticles.

 \par
The paper is constructed as follows. In Section II I formulate the
problem mathematically and carry out the linearization procedure.
Section III is devoted to the procedure of inversing of collision
operator by the use of projection operators method. In Section IV
I derive the exact solution for characteristic diffusion time and
analyze all limiting cases. The classical systems of Lorentz and
Rayleigh gases and their generalization to the quasiparticle
systems are considered in Section V. In Section VI I consider the
generalization of the Kihara approximation on the quasiparticle
quantum systems. The outlines and conclusions are given in Section
VII.

\section{General expressions}

So, consider stationary nonequilibrium state of gaseous mixture of
quasiparticles of two species. The most interesting relaxation
process in such a system is a diffusion \cite{17}, so I will
concentrate on the diffusive processes.

 \par
One of the most essential advantage of the offered theory
undoubtedly is the fact, that it is correct for quantum gases with
any statistics and any dispersion of quasiparticles. All main
outcomes remain applicable for both systems with nonzero chemical
potential, and with chemical potential equal to zero (when the
number of quasiparticles is not conserved). The constructed theory
does not meet principal difficulties in generalization on
multicomponent systems ({\it i.e.} on systems with number of
active components exceeding two ones).

 \par
The evolution of distribution functions $f_k$ (subscript $k=1,2$
numbers the component of the mixture) can be described by the
following set of kinetic equations
 \begin{equation}
 {\bf v}_k\frac{\partial f_k}{\partial{\bf r}}=
 \sum_{j=1,2}C_{kj}(f_k,f_j), \qquad (k=1,2),
 \label{1}
 \end{equation}
where ${\bf v}_k=\partial\epsilon_k /\partial{\bf p}_k$ is the
velocity of quasiparticle of $k$th type, $\epsilon_k$ and ${\bf
p}_k$ are its energy and momentum, respectively; ${\bf r}$ is the
coordinate, $C_{kj}(f_k,f_j)$ is the collision integral, which is
a functional of the distribution functions of mixture components
$1$ and $2$. The particular form of these collision integrals
depends on the concrete physical problem. To find the diffusion
coefficient let us consider the stationary nonequilibrium state of
two--components $1$ and $2$ of the mixture, in which the
quasiparticles number densities are functions of coordinate ${\bf
r}$. In particular, for the gas of thermal excitations such
situation can be realized by creation of constant temperature
gradient.

 \par

Under the considered conditions there are stationary gradients of
partial pressure of components, which result in flow of
quasiparticles. This flow is determined by momentum current
density
 \begin{equation}
 {\bf j}_k=-\sum_{j=1,2} \frac{\rho_k}{\rho}d_{kj}
 \frac{\partial P_j}{\partial{\bf r}},
 \qquad (k=1,2),
 \label{2}
 \end{equation}
where $\rho_k$ is the normal density of $k$th component of the
system, $\rho =\rho_1 +\rho_2$ is the total density, $d_{kj}$ is
the matrix of diffusion times. The partial pressure of
quasiparticles is determined in standard manner \cite{5}
 \begin{equation}
 P_j=\frac{1}{3}\int {\bf p}_j{\bf v}_j f_j d\Gamma_j,
 \label{3}
 \end{equation}
where $d\Gamma_j$ is the measure in phase space. The density of
$k$th component can be written in the universal form \cite{15}
 \begin{equation}
 \rho_k=-\frac{1}{3}\int {\bf p}_k^2 f'_k d\Gamma_k,
 \label{4}
 \end{equation}
where $f'_k=\partial f_k/\partial\epsilon_k$. Relations
(\ref{2})--(\ref{4}) are suitable for the quasiparticles with
arbitrary dispersion law, statistics, and chemical potential (I
mean both for zero chemical potential and for nonzero one). Note,
the definition of normal density (\ref{4}) does not depend
explicitly on such notations as mass or number density.

 \par
To exclude convective transport in a quasiparticle system, {\it
i.e.} to investigate only dissipative processes, hereinafter I
shall consider the sum of partial pressure of different components
of a mixture to be a constant, so put $P=P_1+P_2=\rm{const}$. The
analytical relation between matrix elements of a matrix of
diffusion times $d_{kj}$ and usual diffusion coefficient $D$ of a
binary mixture can be obtained by a direct comparison of
expression (\ref{2}) with definition of diffusion coefficient.
Thus, one can use the ordinary gas-dynamic definition of current
density of $k$th component of a mixture (see {\it e.g.} Ref. 5)
 \begin{equation}
 {\bf j}_k=\int {\bf p}_k f_k d\Gamma_k.
 \label{5}
 \end{equation}
It appears that for various both quantum and classical physical
systems diffusion coefficient can be written in the most general
form \cite{15,18}
 \begin{equation}
 D=u_D^2 \tau_D,
 \label{6}
 \end{equation}
where $u_D$ is the characteristic velocity, whose analytic form
depends on the particular dispersion law and statistics of
quasiparticles, $\tau_D$ is the characteristic diffusion time to
be determined.

 \par
According to the relations (\ref{2}) and (\ref{5}), for deriving
the diffusion coefficient (\ref{6}) it is necessary to solve a
system of the kinetic equations (1). Below for definiteness the
diffusion in a system with conserved number of quasiparticles
(and, therefore nonzero chemical potential) will be considered.
The calculation for case with not conserved number of
quasiparticles can be carried out within the similar framework.

 \par
Since we are interested in the theory within linear response
approximation, let us assume the deviation of distribution
functions $f_k$ from their locally equilibrium values $f_k^{(0)}$
to be small. So, put as usual
 \begin{equation}
 f_k=f_k^{(0)} + \delta f_k, \qquad
 |\delta f_k| \ll f_k^{(0)}.
 \label{7}
 \end{equation}
The small deviation $\delta f_k$ can be conveniently rewritten in
the form
 \begin{equation}
 \delta f_k = - \frac{\partial f_k^{(0)}}{\partial \epsilon_k}
 g_k
 \label{8}
 \end{equation}
with unknown functions $g_k$. Linearizing the system of kinetic
equations (\ref{1}) we come to the system of linear
integro-differential equations for unknown quantities $g_k$, which
determine the degree of system perturbation
 \begin{eqnarray}
 \frac{{\bf v}_k}{n_k}\frac{\partial P_k}{\partial{\bf r}}=
 C_{kk}g_k+C_{kj}(g_k+g_j) \label{9}\\
 (k,j=1,2; k\not=j). \nonumber
 \end{eqnarray}
Here $n_k$ is the number density of $k$th component, $C_{kj}$ are
the linearized collision operators for the collisions within each
component ($k=j$) and for the collisions between different
quasiparticles ($k\not=j$). The acting of this operators on an
arbitrary function of momentum, say $\xi ({\bf p}_{k,j})$, is
determined by the particular form of collision integrals appeared
in equation (\ref{1}). If we deal with the ordinary binary
collision integral with probability density function $W_{kj}({\bf
p}_k{\bf p}_j|{\bf p}'_k{\bf p}'_j)$ then we obtain \cite{19}
 \begin{eqnarray}
 C_{kj}\xi  ({\bf p}_{k,j})= \int W_{kj}({\bf p}_k{\bf
 p}_j|{\bf p}'_k{\bf p}'_j) f_j^{(0)} ({\bf p}_j)\nonumber \\
 \times
 \left\{ 1\pm f_k^{(0)}({\bf p}_k)\right\}^{-1}
 \left\{ 1\pm f_k^{(0)}({\bf p}'_k)\right\}
 \left\{ 1\pm f_j^{(0)}({\bf p}'_j)\right\}\nonumber \\
 \times
 \left[ \xi ({\bf p}'_{k,j})-\xi ({\bf p}_{k,j})\right]
 d\Gamma_j d\Gamma'_k d\Gamma'_j
 \label{10}
 \end{eqnarray}
for $k\not=j$, and
 \begin{eqnarray}
 C_{kk}\xi ({\bf p}_k) = \int W_{kk}({\bf p}_k{\bf
 p}|{\bf p}'_k{\bf p}') f_k^{(0)} ({\bf p})\nonumber \\
 \times
 \left\{ 1 \pm f_k^{(0)}({\bf p}_k)\right\}^{-1}
 \left\{ 1\pm f_k^{(0)}({\bf p}'_k)\right\}
 \left\{ 1\pm f_k^{(0)}({\bf p}')\right\}\nonumber \\
 \times
 \left[ \xi ({\bf p}'_k)+\xi ({\bf p}')-
 \xi ({\bf p}_k)-\xi ({\bf p})\right]
 d\Gamma d\Gamma'_k d\Gamma'
 \label{11}
 \end{eqnarray}
for $k=j$. The plus and minus in eqs. (\ref{10}) and (\ref{11})
signs correspond to bosons and fermions, respectively.

 \par
Note that in fact our general approach allows to account in such a
manner not only binary collisions, but a variety of more specific
types of interaction, such as decay or conversion processes
\cite{20}, creation or annihilation of quasiparticles, interaction
with boundaries and point defects etc.

\section{Inversing of collision operator}

According to the relation (\ref{3}), the gradient of partial
pressure for quasiparticles with nonzero chemical potential can be
written as
 \begin{equation}
 \frac{\partial P_k}{\partial{\bf r}} =n_k
 \left(\frac{\partial \mu_k}{\partial{\bf
 r}}\right)_{T=\rm{const}},
 \label{12}
 \end{equation}
where $\mu_k$ is the chemical potential of $k$th subsystem, $T$ is
the mixture temperature.

 \par
For further calculations it is convenient to present system
(\ref{9}) in the compact matrix form
 \begin{equation}
 \sum_{k=1,2}|\psi_k\rangle\frac{\partial P_k}{\partial{\bf r}}=
 \hat{\cal C}|g\rangle,
 \label{13}
 \end{equation}
where
 \begin{equation}
 |\psi_1\rangle =\Biggl\vert
 \begin{array}{c}
 {\bf v}_1n_1^{-1}\\
 0
 \end{array} \Biggr\rangle, \quad
 |\psi_2\rangle =\Biggl\vert
 \begin{array}{c}
 0\\
 {\bf v}_2n_2^{-1}
 \end{array} \Biggr\rangle, \quad
 |g\rangle =\Biggl\vert
 \begin{array}{c}
 g_1\\
 g_2
 \end{array} \Biggr\rangle,
 \label{14}
 \end{equation}
are the two--component ket-vectors, defined in infinite Hilbert
space to be specified. The collisional operator matrix $\hat{\cal
C}$ can be conveniently written as a sum
 \begin{equation}
 \hat{\cal C}=\hat{J}+\hat{S},
 \label{15}
 \end{equation}
where the operator matrix
 \begin{equation}
 \hat{J}= \left(
 \begin{array}{cc}
 C_{12} & C_{12}\\
 C_{21} & C_{21}
 \end{array} \right)
 \label{16}
 \end{equation}
contains only the quantities corresponding to interactions between
quasiparticles from different subsystems, and
 \begin{equation}
 \hat{S}= \left(
 \begin{array}{cc}
 C_{11} & 0\\
 0 & C_{22}
 \end{array} \right)
 \label{17}
 \end{equation}
corresponds to the relaxation inside each subsystem. To define
completely the Hilbert space I am working in, let us introduce the
scalar product of two--dimensional bra-vector $\langle\zeta |=
\langle\zeta_1 ({\bf p}_1), \zeta_2 ({\bf p}_2)|$ and ket-vector
$|\chi\rangle =\left(\langle\chi | \right)^{\dag}$ in the
following manner
 \begin{equation}
 \langle\zeta |\chi\rangle =
 \sum_{k=1,2}(\zeta_k|\chi_k) =-\sum_{k=1,2}
 \int\zeta_k^*\chi_k \frac{\partial f_k^{(0)}}{\partial\epsilon_k}
 d\Gamma_k,
 \label{18}
 \end{equation}
where $(\zeta_k|$ and $|\chi_k)$ are the corresponding
one--component bra-vector and ket-vector, respectively. It is easy
to verify that with such a choice of scalar product (\ref{18}) the
collision operator $\hat{\cal C}$ becomes hermitian.

 \par
System (\ref{13}) is the system of non-uniform linear integral
equations. According to the general theory of integral equations
the sought solution $|g\rangle$ of system (\ref{13}) must be
orthogonal to the solution of corresponding uniform equations
 \begin{equation}
 \hat{\cal C}|\phi_1\rangle =0.
 \label{19}
 \end{equation}
The normalized solution of equation (\ref{19}) can be written in
the following form
 \begin{equation}
 |\phi_1\rangle =\frac{1}{\sqrt{3\rho}}\Biggl\vert
 \begin{array}{c}
 {\bf p}_1\\
 {\bf p}_2
 \end{array}\Biggr\rangle.
 \label{20}
 \end{equation}
This vector $|\phi_1\rangle$ corresponds to the total momentum of
our two--component quasiparticle system. In general, equation
(\ref{19}) has another solutions, corresponding to conservation of
energy, particle number etc. I account here only the momentum
conservation law because the sought solution can depend only on
the quasiparticle momenta (see initial equations (\ref{13})). It
is convenient to write the formal solution of equation (\ref{13})
so that the orthogonality condition
 \begin{equation}
 \langle g|\phi_1\rangle=0
 \label{21}
 \end{equation}
is contained explicitly in the solution. For this purpose let us
define the projection operator ${\cal P}_n$ onto the subspace
orthogonal to the vector $|\phi_1\rangle$,
 \begin{equation}
 {\cal P}_n=1-{\cal P}_c, \quad
 {\cal P}_c=|\phi_1\rangle\langle\phi_1|.
 \label{22}
 \end{equation}
As a result, the formal solution of equation (\ref{13}) can be
written in the form
 \begin{equation}
 |g\rangle ={\cal P}_n \left( \hat{\cal C}^{-1} \right){\cal P}_n
 \sum_{k=1,2}|\psi_k\rangle \frac{\partial P_k}{\partial{\bf r}}.
 \label{23}
 \end{equation}
Further we must insert the solution (\ref{23}) into the expression
for current density (\ref{5}) keeping in mind the relations
(\ref{7}) and (\ref{8}). Comparing the obtained result with
definition (\ref{2}) for a matrix of diffusion times we come to
 \begin{equation}
 d_{11}=\frac{\rho_2}{\rho_1}\tau_D, \quad
 d_{12}=d_{21}=\tau_D, \quad
 d_{22}=\frac{\rho_1}{\rho_2}\tau_D,
 \label{24}
 \end{equation}
where
 \begin{equation}
 \tau_D=-\langle \phi_2| \hat{\cal C}^{-1}| \phi_2 \rangle
 \label{25}
 \end{equation}
is the characteristic diffusion time, and
 \begin{equation}
 | \phi_2 \rangle =\frac{1}{\sqrt{3\rho\rho_1\rho_2}}
 \Biggl\vert \begin{array}{c}
 \rho_2 {\bf p}_1\\
 -\rho_1{\bf p}_2
 \end{array} \Biggr\rangle
 \label{26}
 \end{equation}
is the characteristic diffusion vector, which is orthogonal to
$|\phi_1 \rangle$.

 \par
Now the problem is reduced to calculation of the matrix element
(\ref{25}), which contains the inverse matrix operator determined
by integral collision operators.

\section{Exact and limiting expressions for the diffusion
time}

To derive an exact analytical expression for the unknown quantity
(\ref{25}) it is necessary to introduce a full system of
orthonormal two-dimensional vectors $|\phi_n\rangle$ (here
$n=1,2,3\dots$) belong to the infinite-dimensional Hilbert space
with a scalar product (\ref{18}). The concrete choice of a system
of basis vectors in many respects depends on convenience of
calculations within the framework of a concrete physical problem
(see, {\it e.g.} Refs. 5 and 8). In our problem it is convenient
to take vector (\ref{20}) as the first of them, and (\ref{26}) as
the second. The remaining vectors can be arbitrary (for example,
such vectors can be built on the basis of Sonin polynomials in
classical case \cite{5}, or Akhiezer-Aleksin-Khodusov polynomials
\cite{8} in quantum systems), but should satisfy to the
completeness and orthogonality conditions
 \begin{equation}
 \sum_{m=1}^{\infty}|\phi_m\rangle\langle\phi_m|=1, \quad
 \langle\phi_m|\phi_n\rangle =\delta_{mn}.
 \label{27}
 \end{equation}
In a constructed full system of vectors an exact expression for
the diffusion time
 \begin{equation}
 \tau_D=-\left\{ \left( \hat{J}+\hat{S} \right)^{-1} \right\}_{22}
 \label{28}
 \end{equation}
can be reduced to the visual analytical form allowing simple
physical interpretation and providing a possibility to carry out
the in-depth qualitative analysis of the obtained result in
various limiting cases, corresponding to different mechanisms of
equilibration in the system of quasiparticles:
 \begin{equation}
 \tau_D=- \left\{ I_{22} - \sum_{n,m=3}^{\infty}I_{2n}
 \left[ \left( {\cal I}+{\cal S} \right)^{-1} \right]_{nm}
 I_{m2}\right\}^{-1}.
 \label{29}
 \end{equation}
Here the square matrices ${\cal I}$ and ${\cal S}$ contains the
following matrix elements
 \begin{eqnarray}
 \| {\cal I}\|_{nm}
 =I_{nm}=\langle\phi_n|\hat{J}|\phi_m\rangle,
 \label{30}\\
 \| {\cal S}\|_{nm}
 =I_{nm}=\langle\phi_n|\hat{S}|\phi_m\rangle.
 \label{31}
 \end{eqnarray}
To obtain the formal result (\ref{29}) I have used the relations
(\ref{19}) and (\ref{27}).

 \par
The matrices (\ref{30}) and (\ref{31}) are infinite-dimensional
and non-diagonal. Therefore the exact solution (\ref{29}) does not
allow to obtain the closed analytical expression for
characteristic diffusion time. However it is necessary to
emphasize, that the result (\ref{29}) contains explicitly not only
the quantities responsible for interaction between quasiparticles
of different types, but also the matrix elements appropriate to
collisions between the quasiparticles within each subsystem. It
allows to investigate various limiting cases, to find minimum and
maximum values of diffusion time, to construct correct
interpolation formulae and useful analytical models, to make
calculations on computers for concrete physical systems.

 \par
To carry out the detailed analysis of the formal solution
(\ref{29}), it is convenient to concretize slightly the vector
basis $|\phi_n\rangle$ for $n>2$. Namely, let us choose the
remaining vectors in the following form:
 \begin{equation}
 |\phi_{2\alpha+1}\rangle=
 \frac{1}{{\cal N}^{(\alpha)}_1}
 \Biggl\vert
 \begin{array}{c}
 F^{(\alpha)}({\bf p}_1)\\
 0
 \end{array}
 \Biggr\rangle ,
 \label{32}
 \end{equation}

 \begin{equation}
 |\phi_{2\alpha+2}\rangle=
 \frac{1}{{\cal N}^{(\alpha)}_2}
 \Biggl\vert
 \begin{array}{c}
 0\\
 F^{(\alpha)}({\bf p}_2)
 \end{array}
 \Biggr\rangle,
 \label{33}
 \end{equation}
where $\alpha =1,2,3,\dots$, $F^{(\alpha)}({\bf p}_k)$ is the
system of properly chosen orthogonal polynomials with the norm
${\cal N}^{(\alpha)}_k=(F^{(\alpha)}({\bf p}_k)|F^{(\alpha)}({\bf
p}_k))^{1/2}$, such that $(F^{(\alpha)}({\bf p}_k)|{\bf p_k})=0$
(it is clear that such polynomials cannot contain first degree of
momentum). Such a choice (\ref{32}), (\ref{33}) provide a
possibility to "separate" in some manner the component of the
mixture. The matrix ${\cal J}+ {\cal S}$ takes now the following
form
 \begin{eqnarray}
 {\cal J}+ {\cal S}= - \left(
 \begin{array}{cccc}
 \nu_{1}^{(11)}&\nu_{12}^{(11)}&\nu_{1}^{(12)}&\ldots \\
 \nu_{21}^{(11)}&\nu_{2}^{(11)}&\nu_{21}^{(12)}&\ldots \\
 \nu_{1}^{(21)}&\nu_{12}^{(21)}&\nu_{1}^{(22)}&\ldots \\
 \vdots & \vdots & \vdots & \ddots
 \end{array}
 \right) \nonumber \\ - \left(
 \begin{array}{cccc}
 \omega_{1}^{(11)} & 0 & \omega_{1}^{(12)} & \ldots \\
 0 & \omega_{2}^{(11)} & 0 &\ldots \\
 \omega_{1}^{(21)} & 0 & \omega_{1}^{(22)} & \ldots \\
 \vdots & \vdots & \vdots & \ddots
 \end{array}
 \right),
 \label{34}
 \end{eqnarray}
where I have introduced the following "higher" interaction rates:
 \begin{equation}
 \nu_{jk}^{(\alpha\beta)}=
 -\frac{(F^{(\alpha)}({\bf p}_j)|C_{jk}
 |F^{(\beta)}({\bf p}_k))}{{\cal N}_j^{(\alpha)}
 {\cal N}_k^{(\beta)}}
 \label{35}
 \end{equation}
and
 \begin{equation}
 \nu_{j}^{(\alpha\beta)}=
 -\frac{(F^{(\alpha)}({\bf p}_j)|C_{jk}
 |F^{(\beta)}({\bf p}_j))}{{\cal N}_j^{(\alpha)}
 {\cal N}_j^{(\beta)}}
 \label{36}
 \end{equation}
 for collisions of different quasiparticles, and
 \begin{equation}
 \omega_{j}^{(\alpha\beta)}=
 -\frac{(F^{(\alpha)}({\bf p}_j)|C_{jj}
 |F^{(\beta)}({\bf p}_j))}{{\cal N}_j^{(\alpha)}
 {\cal N}_j^{(\beta)}}
 \label{37}
 \end{equation}
corresponding to relaxation within each subsystem. The
representation (\ref{34}) helps us to understand the explicit
structure of the formal solution (\ref{29}). This particularly
provide a possibility to study the most important limiting cases
and to reveal their physical matter.

 \par
So, in case of infinitely fast establishment of an equilibrium
between quasiparticles of identical type (so-called complete
control regime \cite{14,21}), when the strong inequalities
 \begin{equation}
 \omega_j^{(\alpha\beta)}\gg\nu_j^{(\alpha\beta)},
 \nu_{jk}^{(\alpha\beta)},
 \quad (j,k=1,2)
 \label{38}
 \end{equation}
take a place, the second term in the brackets of general result
(\ref{29}) vanishes and diffusion time (\ref{28}) is given by the
following simple formula \cite{19}
 \begin{equation}
 \tau_D^{(\rm{cc})}\equiv\tau_D^{(\rm{min})}=
 -\frac{1}{I_{22}}= \left( {\tau_{12}^{(0)}}^{-1}+
 {\tau_{21}^{(0)}}^{-1}\right)^{-1},
 \label{39}
 \end{equation}
where
 \begin{equation}
 \tau_{kj}^{(0)}=-\langle C_{kj} \rangle_k^{-1}, \quad
 (k,j=1,2; k\not=j).
 \label{40}
 \end{equation}
Here and thereafter the brackets $\langle\dots\rangle$ stand for a
normalized average defined by the relation
 \begin{equation}
 \langle{\cal A}\rangle_k =\frac{1}{3\rho_k}
 ({\bf p}_k|{\cal A}|{\bf p}_k), \quad
 (k=1,2).
 \label{41}
 \end{equation}
According to the momentum conservation law $C_{kj}|{\bf p}_k)
=-C_{jk}|{\bf p}_j)$ the relation between "basic" interaction
rates in equation (\ref{39}) can be found as follows
 \begin{equation}
 {\tau_{12}^{(0)}}^{-1}=\frac{\rho_2}{\rho_1}{\tau_{21}^{(0)}}^{-1}.
 \label{42}
 \end{equation}
Proceeding from hermiticity and negativity of operators $\hat{S}$
and $\hat{J}$ with the help of well--known Cauchy-Bunyakovsky
inequality it is possible to show \cite{19}, that the following
inequality is always valid
 \begin{equation}
 \tau_D\geq\tau_D^{(\rm{min})},
 \label{43}
 \end{equation}
where $\tau_D^{(\rm{min})}$ is determined by eq. (\ref{39})

 \par
In the opposite limiting case of extremely slow establishment of
equilibrium between identical quasiparticles, {\it i.e.} when
 \begin{equation}
 \omega_j^{(\alpha\beta)}\ll\nu_j^{(\alpha\beta)},
 \nu_{jk}^{(\alpha\beta)},
 \quad (j,k=1,2),
 \label{44}
 \end{equation}
the second matrix in eq. (\ref{34}) can be neglected and the whole
formula (\ref{29}) can be converted so that the diffusion time
$\tau_D$ runs up to its maximum value
 \begin{equation}
 \tau_D^{(\rm{max})}= - \langle\phi_2| \hat{J}^{-1}|\phi_2\rangle.
 \label{45}
 \end{equation}
Comparing the main limiting results (\ref{39}) and (\ref{45}) we
come to the important conclusion: The main qualitative difference
between expressions for diffusion time in case of a fast and slow
relaxation inside each subsystem of a mixture consists in a method
of averaging of operator matrices corresponding to interaction
between {\it different} subsystems.

 \par
Very interesting situation can be realized when one, say first,
subsystem equilibrates very slowly $C_{11}\to 0$, but relaxation
in another one is extremely fast $C_{22}\to\infty$, so that
 \begin{equation}
 \omega_1^{(\alpha\beta)}\ll\nu_1^{(\alpha\beta)},
 \nu_{12}^{(\alpha\beta)}, \quad
 \omega_2^{(\alpha\beta)}\gg\nu_2^{(\alpha\beta)},
 \nu_{21}^{(\alpha\beta)}.
 \label{46}
 \end{equation}
In this case the results (\ref{29}) and (\ref{34}) can be
converted as follows
 \begin{equation}
 \tau_D \approx -\langle C_{12}^{-1}\rangle_1 -
 \langle C_{21} \rangle_2^{-1}.
 \label{47}
 \end{equation}
Let us pay attention to the principal difference between first and
second terms in relation (\ref{47}). Again, in the first term we
average the time, while in the second one we average the rate of
interaction between quasiparticles of different species. According
to the Cauchy-Bunyakovsky inequality the characteristic time
$\langle C_{jk} \rangle^{-1}$ is always less than the time
$\langle C_{jk}^{-1} \rangle$ for any momentum dependence of
$C_{jk}$. If the collision operator $C_{jk}$ does not depend on
the quasiparticles momenta, then $\tau_D\equiv
\tau_D^{(\mbox{min})}$. If $C_{jk}$ depends on momenta, then speed
of equilibration of a system depends on how fast the equilibration
between identical quasiparticles is \cite{15,19}. Such situation
brightly manifests in a phonon--impuriton system of superfluid
mixtures of helium isotopes \cite{22} and in phonon systems in
solids \cite{16}. There is a two--stage mechanism of relaxation in
these systems. At the first stage quasiparticles of the second
type interact only with those quasiparticles of first type, whose
momenta correspond to maximum of collision operator $C_{12}$. At
the second stage establishment of an equilibrium in a system is
determined by the interaction of quasiparticles of $1$st type with
each other. This occurs because the quasiparticles of first type
with minimum $C_{12}$ prefer to interact with such quasiparticles
within their subsystem, which are already at equilibrium with
second subsystem.

 \par
The significance of manner of averaging of collision operators can
be illustrated on the simple example of lattice thermal
conductivity in solids. The interaction rate between phonon and
scattering center $\nu_{phi}\propto C_{phi}$ (say, point defect or
impurity) is proportional to $p_{ph}^4$ (so-called Rayleigh
scattering). So, the integral $\langle C_{phi}^{-1} \rangle_{ph}$
diverges at zero momentum and the corresponding relaxation time
tends to zero. This means that nonequilibrium long wavelength
phonon simply "does not see" an impurity. At the same time the
quantity $\langle C_{phi} \rangle_{ph}$ is finite and leads to
finite thermal current. So, the mechanism of equilibration of such
system can be drawn as follows: at first all phonons come to
quasi-equilibrium in their own subsystems, which corresponds to
some stationary flux of phonons, then they begin to scatter on the
impurities and point defects. As it can be seen from the results
(\ref{29}), (\ref{34}), (\ref{39}), (\ref{45}), and (\ref{47}),
analogous competition mechanisms can occur in any two--component
quasiparticle system.

 \par
Calculation of $\langle C_{jk} \rangle_{j}$ for particular
physical systems does not meet any difficulties. However, to
calculate $\langle C_{jk}^{-1} \rangle_{j}$ we must inverse the
collision operator $C_{jk}$, which is not a straightforward
operation. As a rule, it can be done by replacing $C_{jk}$ with
some characteristic rate of interaction between quasiparticles
$\nu_{jk}=\nu_{jk}({\bf p}_k)$, and then by straightforward
averaging the value $\nu_{jk}({\bf p}_k)^{-1}$, which is simply a
multiplying operator. In the next Section I consider the most
popular models in various problems associated with two-component
classical gases or with condensed media whose transport properties
are determined by the processes in two-component quasiparticle
systems.

\section{The Rayleigh and Lorentz models for two-component gaseous
mixtures}

Any theory claiming for a solution of some complicated special
problems, should first of all be agreed necessarily with some
fundamental results in the most simple limiting cases. In the
kinetic theory of classical gases the diffusion in Lorentz gas
(diffusion of light very rarefied component in a gas of massive
slow particles) and Rayleigh gas (diffusion of massive particles
with small concentration in a light gas) traditionally are
considered.

 \par
Both classical models, {\it i.e.} the Rayleigh and Lorentz gases
correspond in fact to the limiting case (\ref{46}), but are more
restricted in particles characteristics. Let us start with Lorentz
model. This is the mixture of light component with very small
concentration and the massive gas (buffer component). Let the
light component be $1$st one. Let me briefly review the classical
approach to a problem of diffusion in such a mixture \cite{5,23}.
The strong inequality
 \begin{equation}
 n_1/n_2 \ll 1
 \label{48}
 \end{equation}
was required for light particles to interact only with a massive
component, but not with one another. Large difference in masses of
particles
 \begin{equation}
 m_1/m_2 \ll 1
 \label{49}
 \end{equation}
ensured an elasticity of scattering of particles of a light
component on massive particles and large difference in thermal
velocities of particles of different types. Thus, particles of the
massive component can be treated as fixed and described by their
equilibrium distribution function. With the purpose of calculation
of a diffusion coefficient in such a system the strong
inequalities (\ref{48}) and (\ref{49}) are usually used for
simplification of initial kinetic equation, which can be reduced
to the so-called Boltzmann-Lorentz equation \cite{5,23}. The
consequent solution of this equation gives explicit expression for
a diffusion coefficient.

 \par
Let us show, that the classical results for diffusion in Lorentz
gas can be directly obtained from the general results given by
eqs. (\ref{29}) and (\ref{34}). The operations below have certain
value itself as a generalization of Rayleigh and Lorentz models on
quantum gases of quasiparticles. Such generalization is not
trivial because of impossibility to introduce the mass and
conserved number density for some quasiparticles. So, we will
formulate the problem in terms of normal densities. In this regard
for Lorentz gas we replace two strong inequalities with the
following
 \begin{equation}
 \rho_1/\rho_2\ll 1.
 \label{50}
 \end{equation}
Then we can rewrite vector $|\phi_2\rangle$
 \begin{equation}
 | \phi_2 \rangle =\frac{1}{\sqrt{3\rho_1}}
 \Biggl\vert \begin{array}{c}
 {\bf p}_1\\
 0
  \end{array} \Biggr\rangle \equiv
 \Biggl\vert \begin{array}{c}
 \chi_1\\
 0
  \end{array} \Biggr\rangle.
 \label{51}
 \end{equation}
An equilibrium in massive component leads to the following simple
expression for operator matrix $\hat{J}$ given in general by eq.
(\ref{16}):
 \begin{equation}
 \hat{J}= \left(
 \begin{array}{cc}
 C_{12} & 0\\
 0 & 0
 \end{array} \right).
 \label{52}
 \end{equation}
Then, using the explicit expression (\ref{34}) we can reduce the
general result (\ref{29}) to the following form
 \begin{eqnarray}
 \tau_D^{(L)}=\Biggl\{(\chi_1|C_{12}|\chi_1)-
 \frac{1}{{\cal N}_1^{(\alpha)}{\cal N}_1^{(\beta)}}
 \nonumber\\
 \times
 \sum_{\alpha,\beta =1}^{\infty} \left(\chi_1|C_{12}|
 F^{(\alpha)}({\bf p}_1)\right)
 \left(\left\| (\phi_{2\alpha+1}|C_{12}|\phi_{2\beta+1})
 \right\|^{-1}\right)_{\alpha\beta}
 \nonumber\\
 \times
 \left(F^{(\beta)}({\bf p}_1)|C_{12}|\chi_1\right)
 \Biggr\}^{-1}.
 \label{53}
 \end{eqnarray}
Vectors $|\phi_{2\alpha+1})$ represent a complete set of
orthonormal vector-functions in momentum space of first component.
This allows us to rewrite relation (\ref{53}) as follows
 \begin{equation}
 \tau_D^{(L)}=-\frac{1}{3\rho_1}
 ({\bf p}_1|C_{12}^{-1}|{\bf p}_1)\equiv
 \tau_{12}^{(\infty)}
 \label{54}
 \end{equation}
Further simplification can be achieved if we believe the massive
particles to be fixed during a collision, so that
 \begin{equation}
 v=|{\bf v}_1 - {\bf v}_2|\approx |{\bf v}_1|.
 \label{55}
 \end{equation}
This particularly means that differential cross section depends on
momenta of light particles only. Moreover, the momentum of a light
particle can change only its direction but not an absolute value.
With these assumptions we can write collision operator $C_{12}$ as
 \begin{equation}
 C_{12}=-|{\bf v}_1| \sigma_t n_2,
 \label{56}
 \end{equation}
which is simply a multiplication operator and, therefore, it can
be inversed without any difficulties. Here I introduced the
transport cross section \cite{5}
 \begin{equation}
 \sigma_t=\int{(1-\cos{\theta_{12}})d\sigma},
 \label{57}
 \end{equation}
where $\theta_{12}$ is the scattering angle, $d\sigma$ is the
ordinary differential cross section. As a result, using the
relations (\ref{55})--(\ref{57}) and defining the diffusion
coefficient in binary gas in usual manner \cite{5,23} we come to
the classical result
 \begin{equation}
 D_{12}^{(L)}=\frac{1}{3nn_1}
 \int{f_1^{(0)}\frac{v_1}{\sigma_t}d\Gamma_1}.
 \label{58}
 \end{equation}
The obtained expression (\ref{58}) coincides with analogous
formulae obtained in Refs. 5 and 23 by simplification of the
initial kinetic equation.

 \par
Note that for deriving the formula (\ref{58}) in the frame of
developed here approach it is enough to require fulfilment of
strong inequality (\ref{50}) for normal densities of components of
a mixture, absence of equilibrium in first component, equilibrium
in second component and elasticity of quasiparticles scattering.

 \par
Now I shall consider one more classical example, namely a
diffusion in so-called Rayleigh gas \cite{24}, that is diffusion
of very rarefied massive gas in the light buffer component with
large concentration. In other words, we, as well as in the
previous case, have a mixture of light and massive components, but
under opposite conditions. So we assume the concentration of a
massive component to be so small that
 \begin{equation}
 \rho_2/\rho_1\ll 1.
 \label{59}
 \end{equation}
Now I assume, that the light component of a mixture is already in
equilibrium, while the particles of the massive component almost
do not interact with one another. In this limiting case we again
can use the result (\ref{53}) with a replacement of subscripts
$1\leftrightarrow 2$. In this case, however, the relative velocity
of particles from different components again is determined by a
velocity of light particle. Therefore, the appropriate collision
operator does not depend on momenta, so that in view of an
orthogonality of vectors of selected basis its nongiagonal
elements vanish, {\it i.e.}
 \begin{equation}
 \left(\chi_2|C_{21}|F^{(\beta)}({\bf p}_2)\right)
 =0,   \qquad (\beta=1,2\dots).
 \label{60}
 \end{equation}
Thus, proceeding from the relation (\ref{53}) with a replacement
of subscripts of components, we come to the relation
 \begin{equation}
 {\tau_D^{(R)}}^{-1}=-\frac{1}{3\rho_2}
 ({\bf p}_2|C_{21}|{\bf p}_2)\equiv \tau_{21}^{(0)^{-1}},
 \label{61}
 \end{equation}
which determines diffusion time in a weak solution of massive
component in an equilibrium light gas. Further, proceeding from a
momentum conservation law, we obtain
 \begin{eqnarray}
 {\tau_D^{(R)}}^{-1}=-\frac{1}{3\rho_2}
 ({\bf p}_1|C_{12}^{-1}| {\bf p}_1)
 \nonumber \\
 =\frac{n_1n_2} {3n\rho_2T}
 \int{f_1^{(0)}p_1^2v_1\sigma_td\Gamma_1}.
 \label{62}
 \end{eqnarray}
Expression (\ref{62}) leads to the well known relation for a
diffusion coefficient of massive particles in equilibrium light
classical gas \cite{2,5}. Let me remark, that in Ref. 5 this
result was obtained by an indirect method with the use of Einstein
relation between diffusion coefficient and mobility. The use of
the method developed here has allowed to obtain the formula
(\ref{62}) without engaging artificial approaches, {\it i.e.}
immediately from a general solution (\ref{29}), (\ref{34}). The
developed approach allows to generalize the result (\ref{62}), as
well as (\ref{58}), on quasiparticle systems with arbitrary
dispersion law and statistics, in particular, on such systems, in
which it is impossible to define notation of a mass in its
classical sense.

\section{Kihara approximation}

In the previous sections I considered only the most physically
interesting limiting cases. However, in practice the following
problem can appear: How to calculate some dissipative coefficient
more precisely in the intermediate case, {\it i.e.} when the
considered limiting situations do not take a place. Of course, the
most straightforward way is to compute it numerically using the
formulae (\ref{29})--(\ref{34}). In that case we are forced to
restrict ourselves with some finite matrices in eq. (\ref{29}).
This is an analog of the Chapman--Cowling approximation in
classical gaseous mixtures. In 1949 Kihara \cite{25} proposed
another approximation, which is in fact simpler than
Chapman--Cowling approximation. He simply proposed to neglect the
nondiagonal integral brackets (see Ref. 2), which is exact for
Maxwell molecules. Unfortunately, this approximation is unproved
until now and it can be partially justified only by experience in
numerical calculations for classical gases. In our theory such an
approximation can be introduced by neglecting all the nondiagonal
matrix elements in (\ref{34}). After this procedure the inverse
matrix $\left( {\cal I}+{\cal S} \right)^{-1}$ takes the form
 \[
 \left( {\cal I}+{\cal S} \right)^{-1}=
 \]

 \begin{equation}
 - \left(
 \begin{array}{cccc}
 t_1^{(11)} & 0 & 0 & \ldots \\
 0 & t_2^{(11)} & 0 & \ldots \\
 0 & 0 & t_1^{(22)} & \ldots \\
  \vdots & \vdots & \vdots & \ddots
 \end{array}
 \right),
 \label{63}
 \end{equation}
where I introduced the diagonal characteristic times
 \begin{equation}
 t_j^{(\alpha\beta)}=
 \left[\nu_{j}^{(\alpha\beta)}+\omega_{j}^{(\alpha\beta)}
 \right]^{-1}.
 \label{64}
 \end{equation}
In view of eqs. (63) and (64) we can rewrite the result (29) in
the following form
\begin{eqnarray}
 \tau_D^{-1}\approx -\langle \phi_2|\hat{\cal C}| \phi_2 \rangle +
 \frac{\langle \phi_2|\hat{\cal C}| \phi_3 \rangle
 \langle \phi_3|\hat{\cal C}| \phi_2 \rangle}
 {\langle \phi_3|\hat{\cal C}| \phi_3 \rangle} \nonumber \\
 +\frac{\langle \phi_2|\hat{\cal C}| \phi_4 \rangle
 \langle \phi_4|\hat{\cal C}| \phi_2 \rangle}
 {\langle \phi_4|\hat{\cal C}| \phi_4 \rangle}+\dots .
 \label{65}
 \end{eqnarray}
The series expansion (\ref{65}) gives the correction to the
limiting result (\ref{39}). In practice we should estimate
numerically an appropriate radius of convergence in series
(\ref{65}) and keep necessary number of terms.

 \par
In general it is clear that the Kihara approximation works well
when the momentum dependence of collision operators is weak
enough. But sometimes the series like (65) have infinite radius of
convergence. This indicates simply that it is better to start with
the opposite limiting formula (45) as a zero order approximation.

\section{Conclusions}

In the present work I present the general theory providing a
possibility to investigate diffusion processes in two-component
gas of quasiparticles with arbitrary statistics and dispersion.
The obtained main equations of the theory are correct for both
systems with conserved and not conserved number of quasiparticles,
which is mathematically expressed in not equality or in equality
of a chemical potential to zero, accordingly. The proposed theory
can be generalized on the cases of classical and quantum gaseous
mixtures with arbitrary number of components.

 \par
To solve the formulated problem I start with the system of kinetic
equations (\ref{1}) driving the evolution of corresponding
distribution functions of components of a mixture. After a
standard procedure of a linearization (\ref{7}), (\ref{9}) of a
kinetic problem, I have chosen the basis in infinite-dimensional
two-parameter Hilbert space with a scalar product (\ref{18})
selected so that operator matrix of collision integrals
(\ref{15})--(\ref{17}) becomes Hermitian. As is known, the inverse
matrix of collision integrals does not exist because of the
moments of collision integrals. However, by projecting on the
nucleus of an integral operator of collisions it is possible to
define somewhat inverse matrix. In the present work this procedure
has been made with the help of introducing of projection operator
(\ref{22}) corresponding to conservation of total momentum of a
quasiparticle system. As an outcome it allowed me to obtain the
general expression (\ref{29}) for characteristic time (see also
eq. (\ref{34})), determining speed of a diffusion relaxation in a
system.

 \par
The obtained general result contains explicitly quantities
responsible for interactions between both quasiparticles from
different components and between identical quasiparticles. It
allows to analyze qualitative difference between mechanisms of
equilibration of a whole system in various limiting cases. So, if
the relaxation inside each component of a system is instantaneous,
the diffusion time is determined by an inverse average of
collision operator, describing interaction between quasiparticles
of different types. In the opposite limiting case, when the
equilibrium in a system occurs infinitely longly, the diffusion
time is equal to an average of the inverse collision operator. The
principal difference between these two limiting results can be
easily understood using an example of phonon thermal conductivity
of solids. The thermal conductivity in this case is simply a
diffusion of phonons in a system of fixed "scatterers"
(impurities, boundaries, defects etc.). If the phonons did not
come yet to equilibrium with one another, the thermal conductivity
is determined by average of inverse frequency of scattering of
phonons on scatterers. In case of a long wavelength phonon, such
frequency is proportional to the fourth degree of momentum
(Rayleigh scattering). Therefore when averaging the integral
simply diverges at zero momentum. That is the long wavelength
phonon simply "does not feel" the scatterer. On the contrary, in
case when the phonons came in an equilibrium with one another, the
magnitude of diffusion time appears to be finite, because now the
frequency is averaged, instead of time. This corresponds to the
so-called two-stage mechanism of a relaxation. At first an
equilibrium in phonon gas appropriate to some stationary phonon
flux is established. Then phonons scatter on scatterers. Just this
process results in a finite heat flux.

 \par
The general results, obtained in the work, give correct
expressions for diffusion coefficients of a light gas in a massive
one (Lorentz gas) (\ref{58}) and a massive gas in a light one
(Rayleigh gas) (\ref{62}). Earlier, these classical results were
obtained only with use of artificial methods based on
simplification of the initial kinetic equation. Furthermore these
two most popular model systems can be easily generalized to
describe quasiparticle systems as well as classical gases.

 \acknowledgments
I am deeply grateful to my colleagues Professor I. N. Adamenko and
Dr. K. {\'E}. Nemchenko, collaboration with which had initiated
solution of the given problem when we were working on the problems
concerning dissipative processes in superfluids. I do appreciate
their contribution to the formulation of the problem and
discussion of the main obtained results.


\begin{references}
\bibitem{17}Note, the diffusion in two--component system of
            quasiparticles is in fact connected with a majority of
            another transport processes, say mobility or
            conductivity. For example, lattice thermal conductivity of
            solids is simply a diffusion of phonons.
\end{references}
\end{document}